\documentclass[12pt]{article}
\usepackage{graphicx}
\usepackage{rotating}
\usepackage{dcolumn}
\usepackage{amssymb}
\usepackage{epsfig}
\usepackage{amsmath,amssymb,amsfonts,textcomp}
\usepackage{color}
\begin{document}
\renewcommand{\thefootnote}{\roman{footnote}}
\alph{footnote}
\setcounter{footnote}{-1}
\addtocounter{footnote} {1}
\begin{titlepage}
{\large\bf\noindent How to take turns: the fly's way to encode and decode rotational information }\\
{\it\noindent Ingrid M. Esteves, Nelson M. Fernandes and Roland K\"oberle}\\
\indexspace
{\small\noindent  DipteraLab, Inst. de F\'{\i}sica de S\~ao Carlos,
University of S\~ao Paulo,\\ 13560-970 S\~ao Carlos, SP, Brasil}\\
{\small\noindent Contact e-mail: rk@if.sc.usp.br}\\
\indexspace
\noindent Running head: Neural code of spike trains.
\indexspace
{\large\bf\noindent ABSTRACT}\\
{\bf 
Sensory systems take continuously varying stimuli as their input  and encode features relevant
for the organism's survival into a sequence of action potentials - spike trains.
The full dynamic range of complex dynamical inputs has to be compressed into a set of discrete spike times and 
the question, facing any sensory system, arises: which features of the stimulus are thereby encoded
and how does the animal decode them to recover its external sensory world?

Here we study this issue for the two motion-sensitive H1 neurons of the fly's optical system, which 
are sensitive to horizontal velocity stimuli, each neuron responding to oppositely pointing  preferred directions.
They constitute an efficient detector for rotations of the fly's body about a vertical axis.
Surprisingly
the spike trains $\rho_B(t)$ generated by an empoverished stimulus $S_B(t)$,
containing just the instants when the of velocity $S(t)$ reverses its 
direction,
convey the same amount of global (Shannon) information as spike trains $\rho(t)$ generated by the complete stimulus $S(t)$.
This amount of information
is just enough to encode the instants of velocity reversal.
Yet this suffices to give the motor system just one, yet vital order: 
go left or  right, turning the H1 neurons into efficient analog-to-digital converters.
Furthermore also probability distributions computed from  $\rho(t)$  and  $\rho_B(t)$  are identical.
Still there are regions in the spike trains following velocity reversals, 80 msec long and 
containing about 3-6 msec long spike intervals, 
where detailed stimulus properties are encoded.
We suggest a decoding scheme - how to reconstruct the stimulus from the spike train, which is fast and
works in real time.
}
\end{titlepage}
{\bf\noindent Introduction}\\
All living organisms rely on the sensory nervous system to quickly and reliably extract relevant
information about a changing external environment.
E.g. the visual system receives  time-continuous optical flow patterns and generates 
a set of discrete identical pulses,  called action  potentials or spikes.
This analog-to-digital conversion
means that only a small fraction of {\em relevant} stimulus properties is
actually encoded in the spike train and a lot of effort has been expended in
finding out just  which and how
\cite{Nemenman:2008,Wright:2002,Keat:2001,Wainwright:1999,Laughlin:1999,Lewen:2001,Laughlin:2001,Vickers:2001,Smyth:2003,Theunissen:2000,Kara:2000}.  
Debates have been going on  for decades over the way and means employed by the organism to effect this
{\em dimensional reduction}\cite{Saleem:2008,sharpee:2007,sharpee:2002,Twer:2001,Roweis:2000,Tenenbaum:2000,Barlow:1961} 
without losing essential information.
Although these studies have revealed a wealth of interesting properties,
a much more direct approach 
would be to investigate the animal's performance to a stimulus, manufactured to  contain only {\em relevant} features. 
Once these have been successfully guessed and appropriately tested,
we need  a decoding  algorithm to  reconstruct the stimulus from the spike train.
It should  work in real time to be  available to the animal,
using only data which  are presumably held in a memory. 

To explore these issues we recorded spikes from the two motion sensitive H1 neurons - one for each compound eye - of the fly {\it Chry\-so\-mya megacephala}. Each H1 neuron is sensitive to horizontally moving stimuli,
being excited by  back-to-front motion and inhibited by the oppositely moving scenery. They 
measure rotational velocities of the fly's body around a vertical axis \cite{Hausen:1981}.
In our experiments the fly sees a rigid pattern moving with horizontal velocity $S(t)$ 
either on a Tektronix monitor or on a large translucent screen - see Methods for details.
From $S(t)$ we manufacture an empoverished version $S_B(t)$, which tells the fly only  which way the 
scenery is rotating: either to the left or to the right.
$S_B(t)$  generates spike trains $\rho_B(t)$, which are indistiguishable from the ones $\rho(t)$ 
generated by $S(t)$, as far as global averages over the whole experiment are concerned.
Yet we discover specific local regions on a time scale of $~50-100$ msec, where the responses $\rho(t)$ and
$\rho_B(t)$ do differ. Thus the encoding and decoding process should occur in a multilayered fashion,
involving several time scales.
\newline

{\bf Generating the same response raster with  boxed stimuli}\\
Consider one H1 neuron and let  its excitatory stimuli be positive ($S(t)> 0$) and inhibitory negative ($S(t) <0 $). 
Instants when the stimulus velocity reverses sign 
correspond to  the zero-crossings of the stimulus $S(t)$.
We now discretize our 
continuous stimulus $S(t)$ and generate a discrete {\em boxed} version  $S_B(t)$,
assigning different values to positive and negative velocities:
\begin{equation}
       S_B(t) = \left\{
        \begin{array}{c} 
              S_0, {\sf if} S(t) \ge 0 \\
              -S_0, {\sf if} S(t)  < 0.
       \end{array}
       \nonumber
        \right. 
        \label{skelstim}
\end{equation}

The constant $S_0$ will be chosen so that $S(t)$ and $S_B(t)$ have the same variance
We emphasize that $S(t)$ and $S_B(t)$ have the same zero-crossings.
 \begin{figure}[tbp]
\includegraphics[,height=12cm,width=12cm]{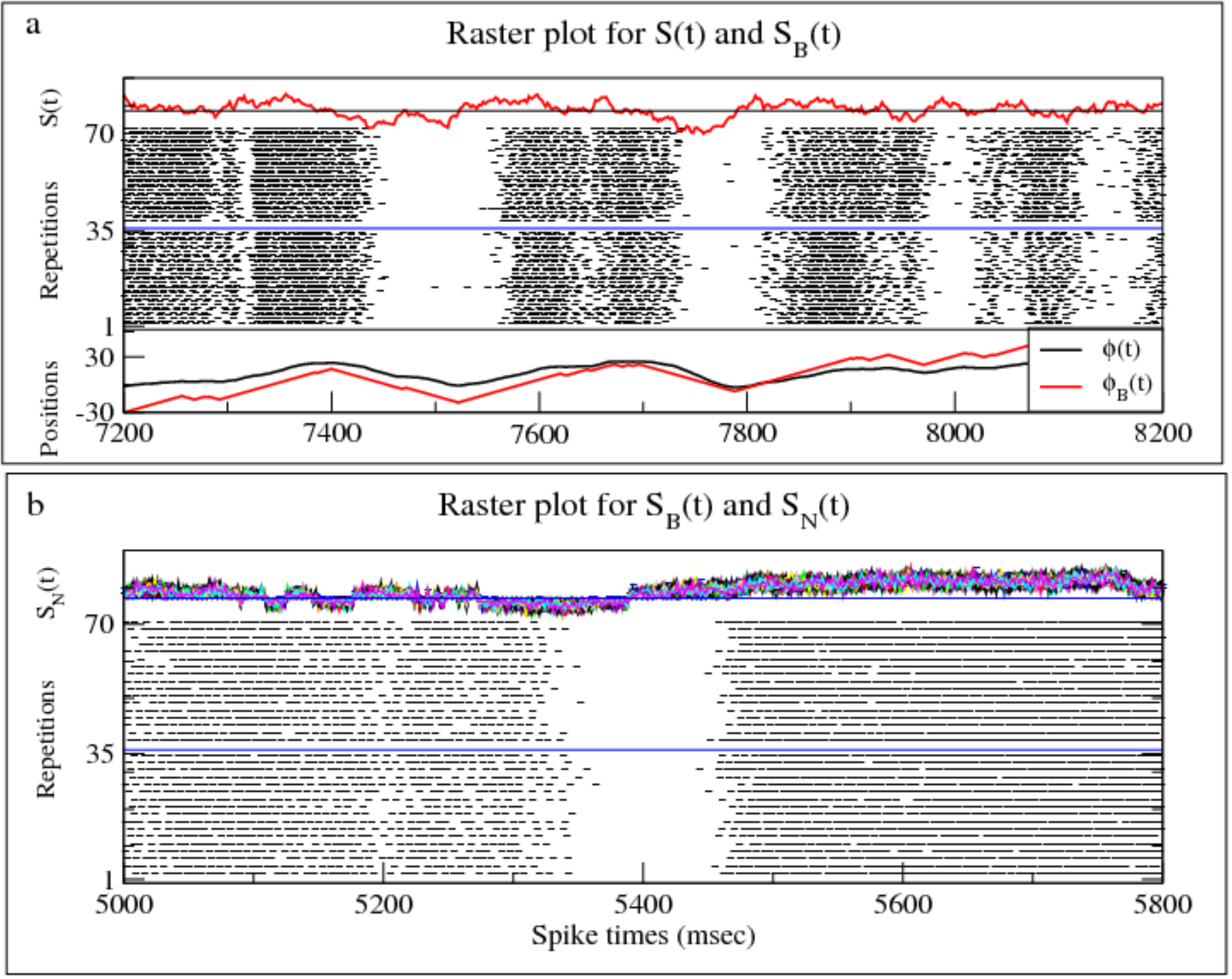}
 	\caption{{\small {\bf Rasterplots of spikes generated by $S(t)$ and its boxed and noisy versions.}
{\bf a}: Section of rasterplots for spikes generated by $S(t)$ and $S_B(t)$.
$S(t)$ is a gaussian velocity signal with correlation time $\tau = 60$ msec, the
raster-time running from $7200$ to $8200$ msec
Repetitions $1:35$ are due to $S(t)$, whereas $36:70$ are due its boxed versions
$S_B(t)$.
The bottom graph shows the actual screen positions
$\phi(t)$ and $\phi_B(t)$ (deg)  as seen by the fly for $S(t)$ and $S_B(t)$. 
Experiment: Tektronix, $\tau = $ 60 msec.
{\bf b}: Section of rasterplots for spikes generated by $S_B(t)$ and noise-added stimuli $S_N(t)$.
Repetitions $1:35$ are due to $S_B(t)$, whereas $36:70$ are due noise-added  gaussian stimuli $S_N(t)$.
On top we show the superposition of all stimuli $S_N(t)$.
Experiment: Natural, $\tau = 200$ msec, $f=0.5$.
   }}
\label{Fig:Raster}
\end{figure}

We  subject the fly alternately to stimuli $S(t)$ and $S_B(t)$, each being  $10$ sec long. 
They are repeatedly shown the fly, the odd/even repetitions being due to $S(t)$/$S_B(t)$ respectively.
From the responses of the H1 neuron we construct a raster, each dot representing a single spike.
In \mbox{Fig.\ref{Fig:Raster}}a we show a section of this raster plot with $35$ repetitions for each stimulus,
where we collected the responses due to $S(t)$ below repetition $35$ and the ones due to $S_B(t)$ above.
The result is completely unexpected: visually
 there is no significant difference between repetitions.
Notice that the plots of the screen positions $\phi(t)$ and $\phi_B(t)$ emphasize, 
that the fly does not even view the
same picture, when subjected to $S(t)$ and $S_B(t)$.
Apparently the fly's H1 sensory system does not pay attention to any  details of stimulus
$S(t)$, which are lost in $S_B(t)$.
                                        
\begin{figure}[tbp]
\includegraphics[height=9cm,width=10cm]{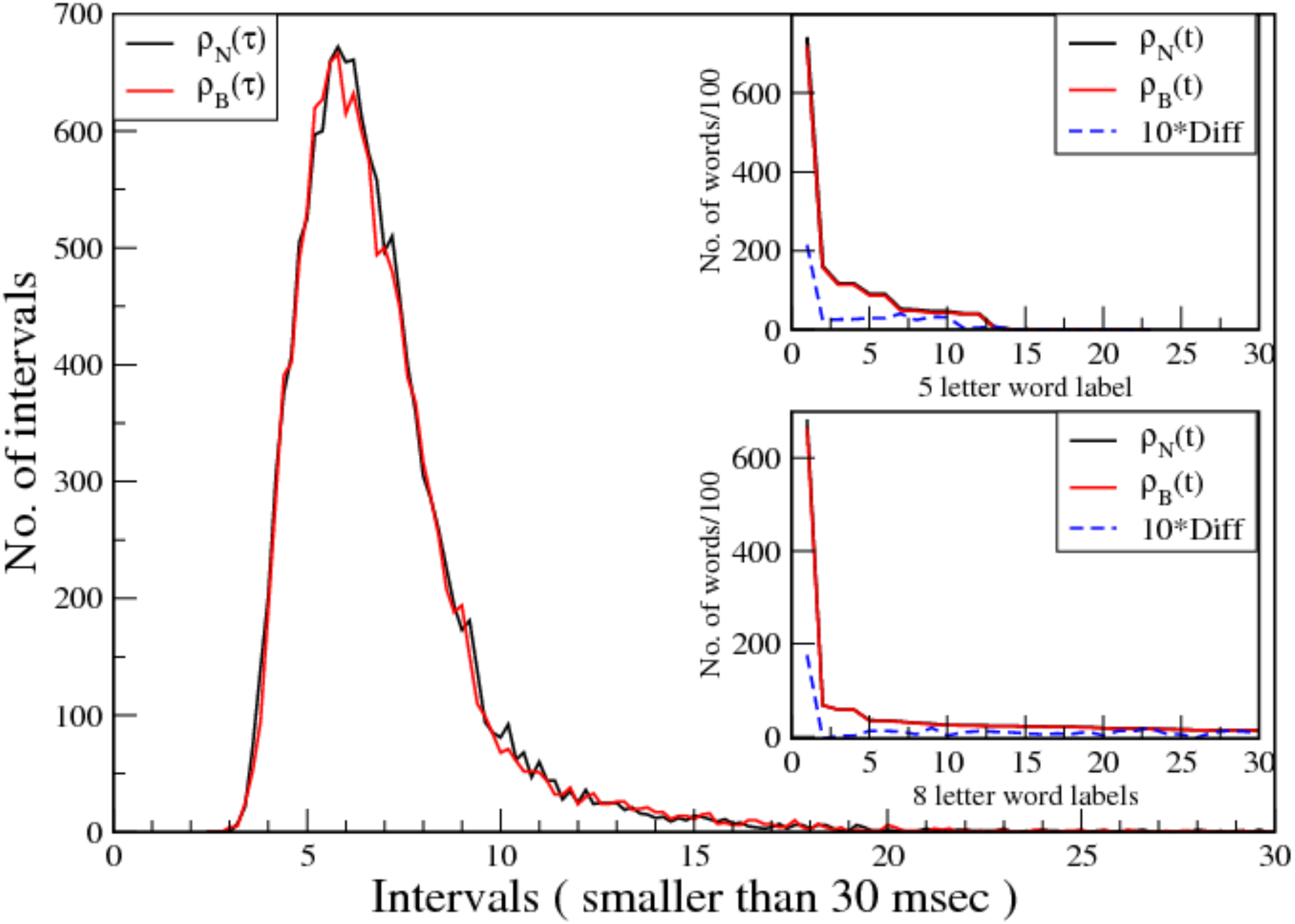}
 \caption{ {\small
{\bf Interval histograms for noise added stimuli.}
Compare the interval histograms generated by stimuli $S_N(t)$ and $S_B(t)$. 
The black (red) curve is the interval histogram of $\rho_N(t)$ ($\rho_B(t)$).
{\bf Upper inset}: Five letter word distributions  for $\rho_N(t) $ and $\rho_B(t)$. 
	The time axis was discretized into $2$ msec wide bins. The presence of a spike
	in each bin was represented by 1 and the absence by 0. Each binary  word is a sequence of five 1's or 0's
	and we have $2^5=32$ possible word labels. We plot the rank-ordered word distribution for $\rho_N(t) $ and $\rho_B(t)$,
	using identical labels for both histograms.
	Their difference, upscaled by $10$, is shown as the dashed (blue) curve. 
{\bf Lower inset}: Same as upper inset for the first thirty eight letter words, but now the word labels 
$W(i)$ and $W_B(i)$ for $\rho_N(t)$  and  $\rho_B(t)$ are not all identical.
E. g. $W(j) = W_B(j), j=1,2,3,4$, but $W(5) = W_B(6), W(6) = W_B(5)$.
Experiment: Tektronix, $\tau = 60$ msec,  $f=0.5$.}
}

\label{Fig:Hist_Noise}
\end{figure}

To further explore this striking fact,
we  generate a whole set of stimuli $S_N(t)$ adding a certain amount of random noise to $S_B(t)$, but always preserving
the zero-crossings\footnote{At instants $t_1$, where the addition of noise would have changed the sign of  $S(t_1)$, 
we replace  $S_N(t_1)$ by  $-S_N(t_1)$.}.
The noise added is measured by the ratio $f \equiv  \sigma(S-S_N)/\sigma(S_N)$, where $\sigma$ is the stimulus variance.
Subjecting the fly to the set $S_B(t), S_N(t)$,
we obtain the raster of \mbox{ Fig.\ref{Fig:Raster}}b. Repetitions $1:35$  are
due to $S_B(t)$ and the the remaining  ones  due to $S_N(t)$, the noise being different for each repetition 
$36:70$.
Again no difference can be seen between the two sets. 
For a quantitative test, we compute the
 interval histograms due to  $\rho_N(t)$ and  $\rho_B(t)$, shown in
 \mbox{Fig.\ref{Fig:Hist_Noise}}: the are the same. 
Additionally all the rank-ordered word distributions of 
 $\rho_N(t)$ and  $\rho_B(t)$  look identical.
In the  insets of Fig.\ref{Fig:Hist_Noise} we plot these distributions for binary words of lengths $L=5,8$.
This astonishing identity also hold for longer words, although the word lables for $L>7$ are then not all identical.

\begin{figure}[tbp]
\includegraphics[height=9cm,width=10cm]{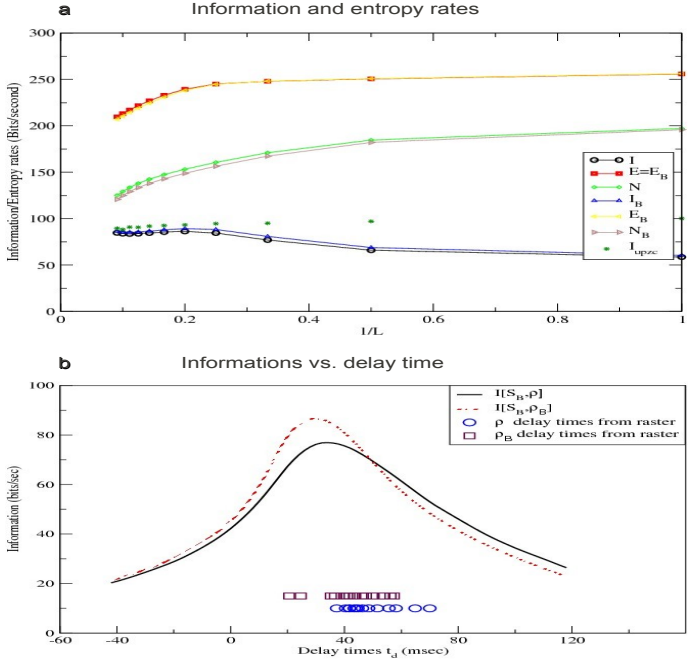} 
 \caption{{\small  {\bf Information and entropy rates. a:} Information and Entropy rates vs. 1/wordlength. 
Spike times were discretized at $2$ msec and the cell's responses were represented as binary words of length $L$ bins.
The word-entropies  $E\equiv E[\rho]$ and $E_B\equiv E[\rho_B]$ are identical. 
$N = E[\rho|S]$ and  $N_B = E[\rho_B|S_B]$
 - the entropies, which the distributions of words convey about the respective stimuli - 
and $I\equiv I[\rho|S]$, $I_B\equiv I[\rho_B|S_B]$ coincide within errors. 
Errors are $\sim 10\%$. $I_{upzc}$ is the information-rate to locate up-zero-crossings.
Experiment: Tektronix, $\tau = 60$ msec. }
{\small {\bf b:} Delay times from $I[S_B,\rho]$, $I[S_B,\rho_B]$ and raster plots. 
Squares, circles are delay times of common ZC's for $\rho$, $\rho_B$ extracted directly from raster plots.
To compute $I[S_B,\rho]$ and $I[S_B,\rho_B]$ time is binned into $2$ msec. To obtain the
required probability distributions
$pr[S_B|\rho_B](t_d)$,  $pr[S_B|\rho](t_d)$  we proceed as follows.
We sample the spike-train distribution $\rho_B(t)$ through a window four bins long, obtaining a set of
binary words $W_{B}$ with their multplicity $m_B$ and their occurrences $t_1$. 
For each word $W_{B}$ we sample  $m_B$  stimulus 
words $W_{S_B}$ of size four, but at instants $t_1-t_d$. Varying $t_d$  we compute $pr[W_{S_B}|W_B](t_d)$
to obtain  $I[S_B,\rho_B]$. We repeat the procedure for $\rho(t)$.
Experiment: Tektronix, $\tau = 160$ msec.}
}
\label{Fig:I2}
\end{figure}

This equivalence holds true for a host of global statistical averages.
Examples, which have been used a lot to
characterize neural responses, are the  Shannon mutual informations (MI)
 $I[\rho|S]$, $I[\rho_B|S_B]$, which 
the spike trains $\rho(t)$,  $\rho_B(t)$  convey about the stimuli $S(t)$,  $S_B(t)$ respectively.
Using the direct method of Ref. \cite{Strong:1998}, we plot in Fig.\ref{Fig:I2}a the
total entropies $E[\rho]$,$E[\rho_B]$ of the spike trains
and the entropies conditioned on their stimuli $E[\rho|S]$,$E[\rho_B|S_B]$. 
Informations are given by
$I[S|\rho] =  E[\rho] - E[\rho|S]$, $I[S_B|\rho_B] =  E[\rho_B] - E[\rho_B|S_B]$. 
Notice that $\rho(t)$ conveys the same amount of MI about the total stimulus $S(t)$ as $\rho_B(t)$ conveys
about the boxed stimulus $S_B(t)$. Thus, as far as the MI is concerned, $\rho(t)$ ignores all the 
features in $S(t)$, which are absent in $S_B(t)$, paying attention only to whether the scenery moves left or right. 
Since the entropy of $S_B(t)$ is about ten times smaller than $S(t)$'s, this amounts to a huge 
entropy reduction with a concomitant increase in coding efficency $e$, defined as the ratio of MI divided by 
stimulus entropy.

The boxed stimulus $S_B(t)$ is characterized by its zero-crossings, half of them being from negative to positive
stimulus values (upcrossings) and the other half the opposite. 
The information to locate all the upcrossings $I_{upzc}$   is about equal to $I[\rho|S]$, 
as shown in Fig.\ref{Fig:I2}a.
It is therefore consistent to assume that one H1 neuron  extracts 
enough information to locate the up-zero-crossings.
Recording from both H1 neurons\cite{Fernandes1:2010}, we obtain an efficiency  $e = 0.42$,
again just enough to locate all (up and down) zero-crossings\footnote{
Under some technical assumptions signals can be reconstructed from their zero-crossings\cite{Logan:1974,Venkatesh:1992}.
} of the stimulus
\footnote{
Since our boxed stimulus $S_B(t)$ has low entropy, it is actually possible to compute the MI as
$I[S_B|\rho_B] =  E[S_B] - E[S_B|\rho_B]$. This yields the same value as the direct method\cite{Strong:1998} and
we also check the symmetry  $I[S_B|\rho_B] = I[\rho_B|S_B]$. As additional bonus we notice that undersampling problems can
be tamed, since we explicitly know the entropy $E[S_B]$ for all word-lengths. 
Values for this entropy obtained from our data can therefore be
corrected. If we assume that the undersampling problems for  $E[S_B]$ and $E[S_B|\rho_B]$ are identical, the same correction
can be applied to the MI $I[S_B|\rho_B]$.
}.
\newline

\begin{figure}[tbp]
\includegraphics[height=10cm,width=12cm]{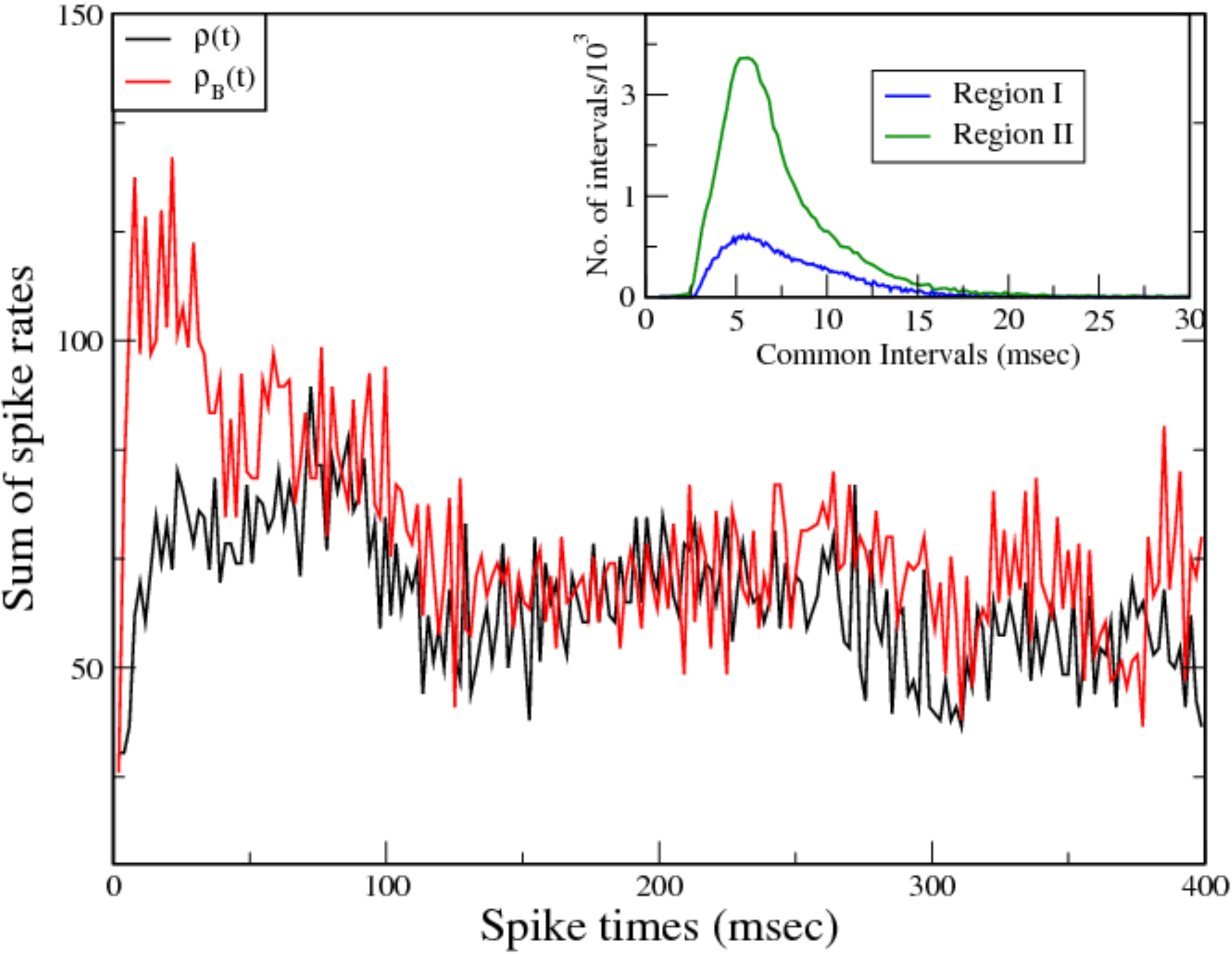}
 \caption{{\small {\bf Spike rates after ZC.}
	Sum of spike rates of $\rho(t)$ (black) and $\rho_B(t)$ (red) summed over  
        $71$  ZC's. The time axis is discretized into $2$ msec bins and
  	 translated so  that  the first spike after ZC occurs at $t=0$. 
        $\rho_B(t)$'s spike rate is always larger and merges with $\rho(t)$'s after $\sim 80 $ msec.
        Experiment: Tektronix, $\tau=400$ msec, $f=0.5$.
	{\bf Inset}: Histograms for intervals ($\leq 30$ msec) common to $\rho(t)$ and $\rho_B(t)$
	for regions I and II, using all experiments.}
}
\label{Fig:rates I/II}
\end{figure}

{\bf\noindent Two coding regions after zero-crossings}\\
Up to now we have endeavored to show that the H1 neuron reponds equally to $S(t)$ and its boxed version,
but there could still be subtle differences, if we look more closely.
To unravel differences between  $\rho(t)$ and $\rho_B(t)$ we select {\em prominent}
zero-crossings (ZC). 
These are instants, when the velocities $S(t)$, $S_B(t)$ change from negative to positive values leading to well defined 
onsets of spiking activity- see Methods for details.
We compute spike rates  after each ZC for a time interval of $400$ msec,
shown in Fig.\ref{Fig:rates I/II}. We observe that 
there are in fact two coding regions I and II, following the first spike after ZC's.
\mbox{Region I\/} spans a time $T_1 \sim 80$ msec. Here
the rate for $\rho_B(t)$ is larger than the  $\rho_B(t)$'s, reflecting the stronger response to
sharper onset of positive stimuli.
This is followed by region II, where the rates for $\rho(t)$ and  $\rho_B(t)$  are the same.

To take a closer look at regions I and II,
we compute in each region intervals common to $\rho(t)$ and $\rho_B(t)$, 
accumulating data ($\sim 50.000$ data points) from ZC's of all experiments.
The histograms of these intervals are shown in the inset of Fig.\ref{Fig:rates I/II}.
Both histograms peak around $6$ msec, but their shape is different.
This could encode the "{\em YES, adjust to turns }" information, which 
the H1 neurons  should extract from the stimulus and relay to the motor system. 
Since the variances of  $S(t)$ and $S_B(t)$ are the same, the spike trains should encode this 
and presumably other ensemble properties in Region II\cite{Fairhall:2001}.
A more detailed investigation of regions I and II could allow the extraction of the precise interval structure associated 
with  the turning command.
Due to the - on average - sharper onset of the boxed stimulus $S_B(t)$ after ZC, the spiking precision of the first and second spike
after all ZC's is always better for 
$\rho_B$ as shown in Table \ref{Table:sp_prec}.   
This underscores the sensibility of the H1 neuron to discontinuities in the stimulus.

\begin{table}[tbp]
     \begin{tabular}{||c|c|c|c|c|c|c ||}\hline 
        $\tau$ (msec) & $\overline{\sigma}^{(1)}$ (msec) &  $\overline{\sigma}_B^{(1)}$ (msec)  &  $\overline{\sigma}^{(2)}$ (msec) & $\overline{\sigma}_B^{(2)}$ (msec)  & $N^{\tau}$ & N$_{B}^{\tau}$      \\ \hline\hline
        10              &  4.35      &  3.37      &   3.63      &  3.34       &  70  &  70       \\ \hline
        30              &  4.97      &  3.83      &   4.81      &  3.73       &  71  &  75        \\ \hline
        80              &  5.81      &  4.19      &   5.75      &  4.60       &  39  &  41       \\ \hline
        100             &  4.66      &  4.34      &   4.67      &  4.56       &  48  &  44       \\ \hline
        200             &  7.23      &  5.42      &   6.43      &  5.53       &  29  &  25       \\ \hline \hline
     \end{tabular}
     \caption{Mean variance of first and second spike after all ZC's for stimuli $S$ and $S_B$ vs. correlation lengths $\tau$.
        $\overline{\sigma}^{(1)}$ ($\overline{\sigma}_B^{(1)}$): Mean variance of 1$^{st}$ spike for $\rho(t)$ ($\rho_B(t)$);  
        $\overline{\sigma}^{(2)}$ ($\overline{\sigma}_B^{(2)}$): Mean variance of 2$^{nd}$ spike for $\rho(t)$ ($\rho_B(t)$).
        For each stimulus $S$ we find all ZC's  and compute the variances $\sigma$ of the 1$^{st}$ and 2$^{nd}$
        spike after each ZC. 
	The mean $\overline{\sigma}$ is taken over the three types of experiments (Natural, Bars, Tektronix), 
	using a total of $N^{\tau}$($N^{\tau}_B$) zero-crossings for $\rho(t)$ ($\rho_B(t)$) respectively. 
        }
     \label{Table:sp_prec}
\end{table}

A further boxed vs. complete stimulus dependent property are the delays
 - the times it takes the fly to generate the 
first spike after ZC. These times exhibit a pronounced dependence on the stimulus history
as shown Fig.\ref{Fig:I2}b. Here we plot delay times, extracted directly from raster plots, for $\rho(t)$
 as circles  and  for $\rho{_B}(t)$ as squares.
For comparision
we also compute {\em mean} delay times $t_{d}$ extracted from  the MI's \mbox{( $I[S_B|\rho], I[S_B|\rho_B] $ ),}
 which the spike trains $\rho(t_1)$ and  $\rho_B(t_1)$  convey about the boxed stimulus $S_{B}(t)$.
When plotted  as a function of  $t_{d}$ they show a pronounced peak, which provides a 
{\em mean-information-delay-time}. 
As we move off the peaks, 
the correlation between stimulus and spike-train vanishes and - although not shown in the figure -
all information rates actually do vanish for large $|t_d|$.
Notice that the peak values are identical to the MI obtained by the direct method of Ref. \cite{Strong:1998}.
In all cases the neuron responds faster by $\sim 10 $ msec to the boxed stimuli, emphasizing again
the systems sensitivity to sharp stimulus variations.
newline
\begin{figure}[tbp]
\includegraphics[height=10cm,width=13cm]{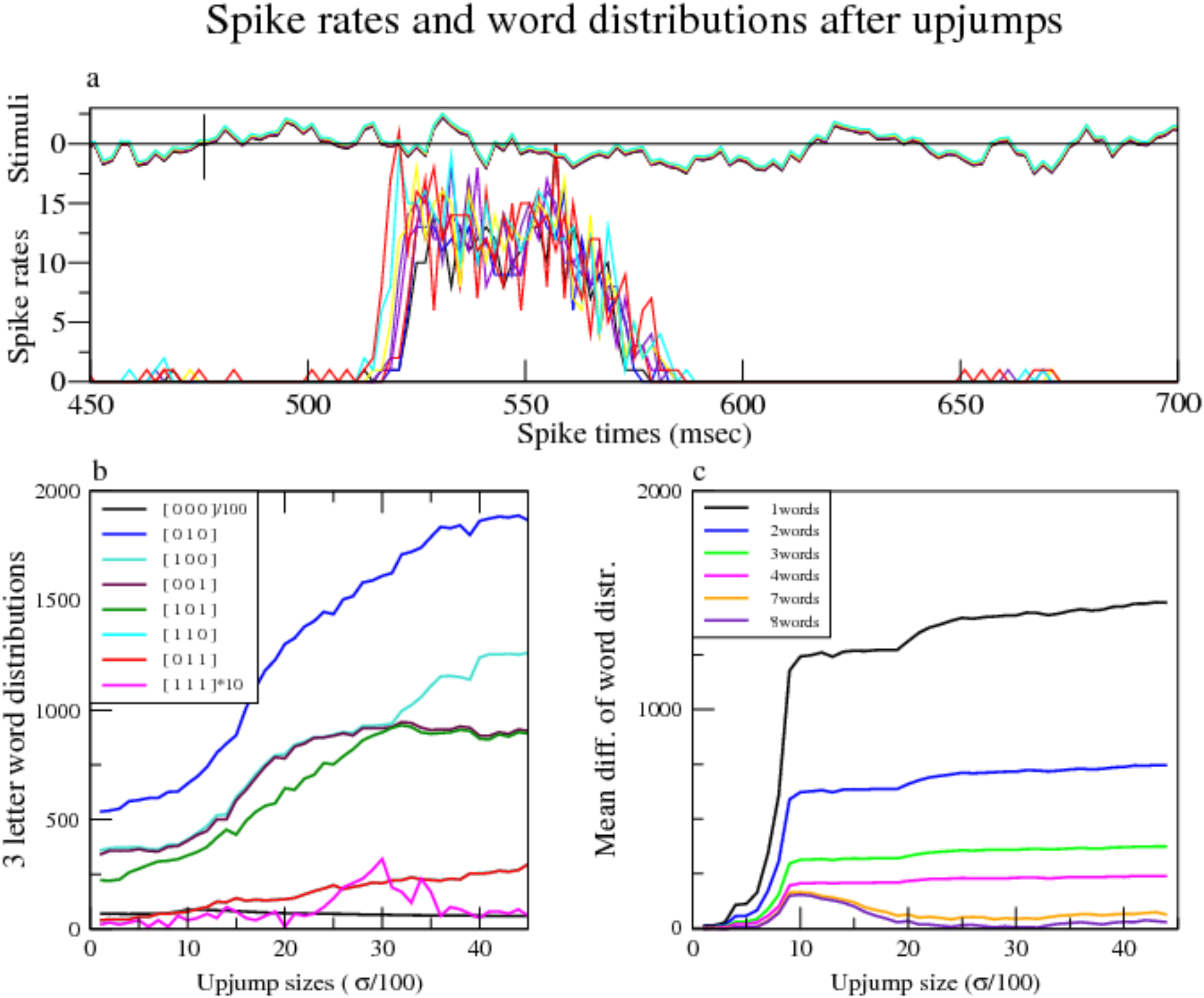} 
\caption{ {\small {\bf a.} 
Spike rates (poststimulus time histograms) after upjump for jumpsizes \mbox{$n*\sigma/100, n=1,2,..,45$.} 
The peak between
$500$ and $600$ msec is due to the zero-crossings at $t\sim 475$ msec - see stimuli on top. 
Its structure clearly depends
on stimulus variance, as revealed by the two plots below.
{\bf b:} 3 letter word distributions $W^{(3)}_i(n)$ in the peak plotted against $n$.
The word distribution for no spikes ($[0 0 0 ]$) is downscaled by a factor of $100$ and the one for 
all occupied by spikes ($[1 1 1]$) is upscaled by a factor of $10$. Binsize is $2$ msec.
{\bf c:} Mean of the word-distribution difference  $W_i(n)-W_i(1)$ for word-lengths $1,2,3,4,7,8$.
Experiment: Tektronix, $\tau =200$ msec.}}
\label{Fig:upjump}
\end{figure}

{\bf\noindent Sensitivity of H1 to temporal discontinuities}\\
If one of the foremost aims of the H1 neurons is the extraction of  temporal
discontinuities from the optical flow, in order to respond preferentially only to these,
the fly has to decide on a {\em threshold}: 
how much stimuli
have to rise above the background for them to be classified as discontinuities?

Let us measure discontinuities by changes in stimulus variance $\delta\sigma$, defining  $f\equiv \delta\sigma/\sigma$.
To study this situation, often stimuli are chosen, which are piecewise constant. 
Since the variance of a constant equals zero, $f=\infty$ and the fly obviously interprets this as a discontinuity, even if $\delta\sigma$ is small.
To simulate a more realistic situation, we expose the fly to a series of random gaussian stimuli $S_n(t)$
with correlation times $\tau = 200$ msec and variances $\sigma_n$. They are generated from a common stimulus $S_0(t)$ with standard deviation
$\sigma_0 = 7$ deg, by upscaling its standard deviation as  $\sigma_n = n*\sigma_0/100, n=1,2,3,..,45$. 
Fig.\ref{Fig:upjump}a shows  the spike rates generated by the stimuli on top. 
The zero-crossing for $n=1$ occurs at $t_1=475$ msec and for $n=45$ at $t_{45} = 478$ msec with intermediate values
for the other n-values.
These zero-crossing initiate the  large spike-rate  at $t \sim 510$ msec.
To expose the structure of this peak's spike patterns,
we compute the  distributions for all 3 letter words $ W^{(3)}_i(n), n=1,2,3,..,45, i=1,2,..,8=2^3$  
in the peak, shown in Fig.\ref{Fig:upjump}b.
The distributions change for upjump sizes $n > 3$.
To emphasize this, we subtract the distributions for the smallest  upjump to get 
 $dW^{(3)}_i(n) = W^{(3)}_i(n) - W^{(3)}_i(1)$ and take 
their mean  $\overline{dW^{(3)}_i} = <dW^{(3)}_i(n)>_n$. In Fig.\ref{Fig:upjump}c 
we show these means for 
word-sizes $1,2,3,4,7,8$,
which clearly show a steep rise starting at the $3 \%$ upjump.
A distributional encoding of the variance can be seen: words of all  lengths contribute to small ($n\leq 15$) upjumps,
whereas large upjumps are encoded into words of length $\leq 4$.
Thus under our experimental conditions the  threshold for  detection is a $\sim 3\%$ change in variance.
\newline

{\bf\noindent Discussion}\\
An efficient representation of the sensory input is one of the main requirements to be 
satisfied by a neural code\cite{Barlow:1961}.
Shannon's mutual information between stimulus and spike train is often used to asses this efficiency.
Our results show that this measure is too coarse, since it is unable to distinguish the complete stimulus from
its boxed version. We find in fact that the mutual information is only sufficient to encode the zero-crossings of
the stimulus.
Yet a closer look reveals a multilayered \mbox{scheme\cite{Fairhall:2001,Baptista:2006}}
of which we observe two layers,
suggesting the following scenario.
In a region  $80$ msec long after spiking onset following zero-crossings we observe spike patterns, about $6$ msec long, which encode turning commands, whereas
stimulus ensemble properties, like stimulus variance, are encoded on time scales longer than $80$ msec.
If variance changes together with zero-crossings, the fly's detection time is much shorter: from the data of 
Fig.\ref{Fig:upjump} we extract a time-scale of $\sim 37$ msec for a change in stimulus variance of $\geq 3\%$.

Once we  know that the spike train carries just enough information to encode the position of all the zero-crossings,
what does this tell us about the decoding problem: 
{\bf how does the fly recover the stimulus from H1's spike train?}
This information allows the extraction of the zero-crossings (or reconstruction of the boxed stimulus $S_B(t)$ up to a scale factor) rather straightforewardly
for stimulus correlation times $\tau>200$ msec:
select a spike train for some repetition in a raster and search it
for  $10$ msec long gaps, followed by 3-4 spikes less than
$\sim 10$ msec apart - this will locate up-zero-crossings $ZC_{sp}$
\footnote{
This recipe is only a rough guide to illustrate a possible path to extract ZC's.
}. 
They coincide with all the up-zero-crossings $ZC_r$ obtained from the spike rate.
Peaks signalling onset of spiking activity and locating $ZC_r$, 
are defined as spike rates larger the 60\% of the repetition number and no preceeding spiking activity for
$20$ msec. 
This definition of  $ZC_r$ 
excludes zero-crossings too small to be relevant for the fly
and extracts only the {\em important} ones.
An identical procedure applied to the responses of the contra-lateral H1 neuron locates the down-zero-crossings.
For correlation times smaller than $200$ msec more elaborate schemes have to be adopted.

Since not all zero-crossings are 
equivalent it stands to reason, that more information is allocated to the {\em important} ones.
It is thus probable, using the spike patterns stored in region I, 
that more than $S_B(t)$ can be recovered from a spike train. 
Since we know that the H1 neuron has acces to the mean and variance of the stimulus\cite{Brenner:2001},
a fast local reconstruction procedure could be accomplished in two (or more) stages.
First  the {\em important} zero-crossings would be extracted from the spike trains, 
the other encoded  properties supplying the scale factor  required 
to reconstruct $S_B(t)$.
In the second stage
more stimulus details could be included using e. g. the usual Volterra series 
reconstruction\cite{Spikes,Fernandes:2009}, although this
requires the knowledge of stimulus-spike correlation functions like  
$<S(t_1)\rho(t_2)>$, whose update at the milisecond time scale is problematic.

We have only  analyzed the output of H1 neurons and suggested a way to extract the specific sequence of spikes,
which would relay the turning command to the motor neurons. 
The fly uses also mechanosensory
organs, gyroscopes using Coriolis forces, to detect fast self-rotations\cite{Pringle:1948,Dickinson:1999}. 
These are the halteres, beating at the same 
frequency as the  wings. The visual and mechanosensory inputs are then fused to obtain a 
more robust estimate of the stimulus\cite{Houston:2009,Fox:2010}.
Therefore we have to keep in mind, that there are many other sensory pathways, which
respond to stimulus details not detected by the H1 neurons.
\newline

{\bf\noindent Methods}\\
{\bf Experimental Setup and Preparation}\\
Immobilized flies 
were shown a rigidly moving 
scenery, while action potentials - spikes - were recorded extracellularly from H1 
neurons.
We used two experimental setups: {\em Tektronix} and {\em Slide}.
In the  Tektronix setup, the fly views  vertical bar patterns on a Tektronix 608 monitor, 
whose update-rate  was  $500$ Hz, so that the bar pattern moved by $\delta x= v(t)$ every $2$ ms\cite{Lirio:2010,Almeida:2011}.
In the {\em Slide} setup the fly views a large  ( 60 cm x 40 cm ) translucent convex cylindrical screen, a slide being projected
on the concave side via a simple optical system containing two mirrors.
These are analogically controlled by linear motors, cannibalized from hard disk controllers. They move 
the slide's projection horizontally across the screen\cite{Esteves:2010}. This setup produces a continously  moving image,
but its performance is limited by  mechanical inertia and fatigue. We therefore use it only for stimuli with correlation time
$\tau \geq 60$ msec.
The slide we use shows either a natural scenery or a random vertical bar pattern - called "Natural" or "Bars" in the text.
Although we recorded simultaneously from both H1 neurons, for simplicity 
we only use  one neuron recordings and
 select a subset of these for our figures. All our conclusion hold for two neuron recordings\cite{Fernandes1:2010}.\\
All experiments were repeated involving about $30$ flies.\\
{\bf Stimulus}\\
The velocity stimuli $v(t)$ were taken from Gaussian distributions with  exponentially decaying correlation 
functions $\sim exp(-t/\tau)$.\\
{\bf Searching for zero crossings}\\
When the stimulus crosses from negative to positive values, one of the H1 neurons typically starts to generate spikes
after a delay time of $\sim 40$ msec. To locate a well defined peak of spiking activity,
we compute the spike rate - or the poststimulus time histogram -
summing for each time bin the occurencies of spikes over all repetitions in the raster.
To locate peaks, we require a spike rate larger than half the number of repetitions in
a small window of a couple of msec.
{\em Prominent} instants  are to be preceeded by no spiking activity for at least $100$ msec and 
we require peaks to  be present in both spike trains $\rho(t)$ and $\rho_B(t)$. 
Additionally no new peak is allowed for the next $400$ msec.\\

{\bf\noindent Acknowledgments}\\
\noindent
We thank I. Zuccoloto, L. O. B. Almeida and J. F. W. Slaets
for help with the experiments. 
NMF was supported by a FAPESP and IME by a CNPq fellowship.
The laboratory was partially funded by FAPESP grant 0203565-4.
We thank Altera Corporation for their University program and Scilab for its excellent software. 

\bibliographystyle{plain}

\begin{thebibliography}{10}

\bibitem{Lirio:2010}
L.~Almeida.,Desenvolvimento de instrumenta\c{c}\~ao eletronica para estudos de
  codifica\c{c}\~oes neurais no duto \'optico em moscas.
 Master's thesis, Univ. of S\~ao Paulo, Brazil, 2010.
{\small Available at:
  http://www.teses.usp.br/teses/disponiveis/76/76132/tde-29032007-105503/en.php. }

\bibitem{Almeida:2011}
L.~O.~B. Almeida, J.~F.~W. Slaets, and R.~K\"oberle.
\newblock {VSI}mg: A high frame rate bitmap based display system for
  neuroscience research.
\newblock {\em Neurocomputing}, 2011.

\bibitem{Baptista:2006}
M.~S. Baptista, C.~Grebogi, and R.~K\"oberle.
\newblock Dynamically multilayered visual system of the multifractal fly.
\newblock {\em Physical Review Letters}, 97:178102--1--178102--4, 2006.

\bibitem{Barlow:1961}
H.~Barlow.
\newblock Possible principles underlying the transfomations of sensory images.
\newblock In W.~Rosenblith, editor, {\em Sensory Communication}, pages
  217--234. MIT Press, Cambridge, MA, 1961.

\bibitem{Brenner:2001}
N.~Brenner, W.~Bialek, and R.~de~Ruyter~van Steveninck.
\newblock Adaptive rescaling maximizes information transmission.
\newblock {\em Neuron}, 26:695--702, 2001.

\bibitem{Dickinson:1999}
M.~H. Dickinson.
\newblock Haltere-mediated equilibrium reflexes of the fruit fly {D}rosophila
  melanogaster.
\newblock {\em Philosophical Transactions of the Royal Society}, 354:903--916,
  1999.

\bibitem{Esteves:2010}
I.~M. Esteves.
\newblock Gerador de estimulos visuais para pesquisar o sistema visual da
  mosca.
\newblock Master's thesis, Univ. of S\~ao Paulo, Brazil, 2010.
\newblock \small Available at:
  www.teses.usp.br/teses/disponiveis/76/76131/tde-04102010-171911/en.php.

\bibitem{Fairhall:2001}
A.~L. Fairhall, G.~D. Lewen, W.~Bialek, and R.~R. van Steveninck.
\newblock Efficiency and ambiguity in an adaptive neural code.
\newblock {\em Nature}, 412(6849):787--792, 2001.

\bibitem{Fernandes1:2010}
N.~M. Fernandes.
\newblock {\em Acuidade visual e codifica\c{c}\~ao neural de mosca Chrysomya
  megacephala}.
\newblock Ph{D} thesis, Univ. of S\~ao Paulo, Brazil, 2010.
\newblock \small Available at:
  www.teses.usp.br/teses/disponiveis/76/76131/tde-25032010-161256/en.php.

\bibitem{Fernandes:2009}
N.~M. Fernandes, B.~D.~L. Pinto, L.~O.~B. Almeida, J.~F.~W. Slaets, and
  R.~K\"oberle.
\newblock Recording from two neurons: second order stimulus reconstruction from
  spike trains.
\newblock {\em Neural Computation}, 186(4):399--407, 2009.

\bibitem{Fox:2010}
Jessica~L. Fox, Adrienne~L. Fairhall, and Thomas~L. Daniel.
\newblock {Encoding properties of haltere neurons enable motion feature
  detection in a biological gyroscope}.
\newblock {\em Proceedings of the National Academy of Sciences},
  107(8):3840--3845, 2010.

\bibitem{Hausen:1981}
K.~Hausen.
\newblock Monokulare und {B}inokulare {B}ewegungsauswertung in der {L}obula
  {P}latte der {F}liege.
\newblock {\em Verh. Dtsch. Zool. Ges.}, pages 49--70, 1981.

\bibitem{Houston:2009}
S.~J. Houston and H.~G. Krapp.
\newblock Nonlinear integration of visual and haltere inputs in fly neck motor
  neurons.
\newblock {\em J. Neuroscience}, 29(42):13097--13105, 2009.

\bibitem{Kara:2000}
P.~Kara, P.~Reinagel, and R.~C. Reid.
\newblock Low response variability in simultaneously recorded retinal, thalamic
  and cortical neurons.
\newblock {\em Neuron}, 27:635--646, 2000.

\bibitem{Keat:2001}
J.~Keat, P.~Reinagel, R.~C. Reid, and M.~Meister.
\newblock Predicting every spike: a model for the responses of visual neurons.
\newblock {\em Neuron}, 30:803--17, 2001.

\bibitem{Laughlin:1999}
S.~Laughlin.
\newblock Visual motion: Dendritic integration makes sense of the world.
\newblock {\em Current Biology}, 9:R15--R17, 1999.

\bibitem{Laughlin:2001}
S.~B. Laughlin.
\newblock Energy as a constraint on the coding and processing of sensory
  information.
\newblock {\em Current opinion in neurobiology}, 11:475--480, 2001.

\bibitem{Lewen:2001}
G.~D. Lewen, W.~Bialek, and R.~R. van Steveninck.
\newblock Neural coding of naturalistic motion stimuli.
\newblock {\em Network-Computation in Neural Systems}, 12(3):317--329, 2001.

\bibitem{Logan:1974}
Jr. Logan.
\newblock Information in the zero crossings of bandpass signals.
\newblock {\em Bell Systems Technical Journal}, 56:487 -- 510, 1974.

\bibitem{Nemenman:2008}
I.~Nemenman, G.~D. Lewen, W.~Bialek, and R.~R. van Steveninck.
\newblock Neural coding of natural stimuli: Information at sub-millisecond
  resolution.
\newblock {\em Plos Computational Biology}, 4:e1000025, 2008.

\bibitem{Pringle:1948}
J.~W.~S. Pringle.
\newblock The gyroscopic mechanism of the halteres of diptera.
\newblock {\em Philosophical Transactions of the Royal Society}, 233:347--385,
  1999.

\bibitem{Spikes}
F.~Rieke, D.~Warland, R.~R. {van Steveninck}, and W.~Bialek.
\newblock {\em Spikes -exploring the neural code}.
\newblock MIT Press, Cambridge, USA, 1997.

\bibitem{Roweis:2000}
S.~T. Roweis and L.~K. Saul.
\newblock Nonlinear dimensionality reduction by locally linear embedding.
\newblock {\em Science}, 290:2323--2229, 2000.

\bibitem{Saleem:2008}
A.~B. Saleem, H.~G. Krapp, and S.~R. Schultz.
\newblock Receptive field characterization by spike-triggered independent
  component analysis.
\newblock {\em Journal of Vision}, 8(13):1--16, 2008.

\bibitem{sharpee:2007}
T.~Sharpee and W.~Bialek.
\newblock Neural decision boundaries for maximal information transmission.
\newblock {\em Plos ONE}, 2(7):646, 2007.

\bibitem{sharpee:2002}
T.~Sharpee, N.C. Rust, and W.~Bialek.
\newblock Analyzing neural responses to natural signals: {M}aximally
  informative dimensions.
\newblock {\em Neural Computation}, 16:223--250, 2002.

\bibitem{Smyth:2003}
D.~Smyth, B.Willmore, G.E.Baker, I.D. Thompson, and J.D. Tolhurst.
\newblock The receptive-field organization of simple cells in primary visual
  cortex of ferrets under natural scene stimulation.
\newblock {\em J. Neurosci.}, 23:4746--4759, 2003.

\bibitem{Strong:1998}
S.~P. Strong, R.~K\"oberle, R.~R. van Steveninck, and W.~Bialek.
\newblock Entropy and information in neural spike trains.
\newblock {\em Physical Review Letters}, 80(1):197--200, 1998.

\bibitem{Tenenbaum:2000}
J.~B. Tenenbaum, V.~da~Silva, and J.~C. Langford.
\newblock A global geometric framework for nonlinear dimensionality reduction.
\newblock {\em Science}, 290:2319--2323, 2000.

\bibitem{Theunissen:2000}
F.~E. Theunissen, K.~Sen, and A.Doupe.
\newblock Spectral-temporal receptive fields of nonlinear auditory neurons
  obtained using natural sounds.
\newblock {\em J. Neurosci.}, 20:2315--2331, 2000.

\bibitem{Venkatesh:1992}
Y.~V. Venkatesh.
\newblock Hermite polynomials for signal reconstruction from zero-crossings.
\newblock {\em IEEE Proceedings-I}, 139:587--, 1992.

\bibitem{Vickers:2001}
N.~J. Vickers, T.~A. Christensen, T.~Baker, and K.~G. Hildebrand.
\newblock Odour-plume dynamics influence the brain's olfactory code.
\newblock {\em Nature}, 410:466--470, 2001.

\bibitem{Twer:2001}
T.~von~der Twer and D.~I.~A. McLeod.
\newblock Optimal nonlinear codes for the perception of natural colours.
\newblock {\em Network: Comput. Neural Syst.}, 12:395--401, 2001.

\bibitem{Wainwright:1999}
M.~Wainwright.
\newblock Visual adaptation as optimal information transmission.
\newblock {\em Vision Research}, 39:3860--3974, 1999.

\bibitem{Wright:2002}
B.~D. Wright, K.~Sen, W.~Bialek, and A.~J. Doupe.
\newblock Spike timing and the coding of naturalistic sounds in a central
  auditory area of songbirds.
\newblock In T.G. Dietterich, S.~Becker, and Z.~Ghahramani, editors, {\em
  Advances in Neural Information Processing}, volume~14, pages 309--316. MIT
  Press, Cambridge, MA, 2002.

\end{thebibliography}

\end{document}